\documentclass{PoS}

\newcommand{\BR}{{\cal B}}
\newcommand{\pp}{\pi^+\pi^-}

\newcommand{\pim}{\pi^-}

\newcommand{\EE}{e^+e^-}

\newcommand{\jpsi}{J/\psi}

\newcommand{\pcpcjpsi}{\pi^+\pi^-J/\psi}

\newcommand{\y}{Y(4260)}
\newcommand{\zc}{Z_c(3900)}
\newcommand{\zcp}{Z_c(4020)}
\newcommand{\zcpp}{Z_c(4025)}
\newcommand{\hc}{h_c}
\newcommand{\pphc}{\pi^+\pi^-\hc}
\newcommand{\psithr}{\psi(4040)}
\newcommand{\psifou}{\psi(4160)}
\newcommand{\psifiv}{\psi(4415)}
\newcommand{\ddb}{D\bar{D}}
\newcommand{\dstbar}{\bar{D}^{*}}

\newcommand{\dst}{D^{*}}

\def\Journal#1#2#3#4{{#1} {\bf #2}, #3 (#4)}
\def\PRL{Phys. Rev. Lett.}
\def\PRD{Phys. Rev. D}

\title{Recent results from BESIII experiment}

\ShortTitle{Recent results from BESIII experiment}

\author{\speaker{Liaoyuan Dong}\thanks{on behalf of the BESIII collaboration}\\
        Institute of High Energy Physics, Beijing 100049, China\\
        E-mail: \email{dongly@ihep.ac.cn}}

\abstract{
In this talk, we present a selection of recent results from BESIII
collaboration, including observation of the charmoniumlike states,
$\zc$, $\zcp$ and $\zcpp$; observation of $e^+e^-\to\gamma X(3872)$; partial wave analysis
of $\jpsi \to \gamma \eta\eta$; measurement of $D^+\to \mu^+\nu$ and $D^0\to
K^-e^+\nu$, $\pi^-e^+\nu$.
The results are based on the data samples collected with the BESIII detector at central-of-mass energies from $3.900$ to $4.420$~GeV, 
and at the energies of $\jpsi$ and $\psi(3770)$ resonances.}

\FullConference{The European Physical Society Conference on High Energy Physics -EPS-HEP2013\\
		18-24 July 2013\\
		Stockholm, Sweden}

\begin{document}

\section{Introduction}

The BESIII experiment at the BEPCII collider has accumulated the world's
largest data samples at $\jpsi$, $\psi(2S)$, and $\psi(3770)$ resonance energies.
Moreover, the BESIII experiment has recently accumulated the data samples at central-of-mass (CM) energies
from $3.900$ to $4.420$~GeV for $XYZ$ physics.
Table~\ref{scan-data} lists the CM energies and the corresponding luminosities of each energy point.
The results in this presentation are based on the data samples in Table1, 
$\jpsi$ data sample of 225 million events, and $\psi(3770)$ data sample of $2.92$~fb$^{-1}$ integrated luminosity.
\begin{table}[htbp]
\caption{The CM energies and luminosities of each data sample.}
\label{scan-data}
{\scriptsize
\begin{tabular}{cccccccccccccc}
\hline
$\sqrt{s}$~(GeV)       & 3.900 & 4.009 & 4.090 & 4.190 & 4.210 & 4.220 & 4.230  & 4.245 & 4.260 & 4.310 & 4.360 & 4.390 & 4.420 \\
${\cal L}$~(pb$^{-1}$) & 52.8  & 482.0 & 51.0  & 43.0  & 54.7  & 54.6  & 1090.0 & 56.0  & 826.8 & 44.9  & 544.5 & 55.1  & 44.7  \\ 
\hline
\end{tabular}
}
\end{table}

\section{Observation of $Z_c(3900)$ in $e^+e^-\to\pi^+\pi^- J/\psi$}

Study of the nature of the $\y$ becomes one of the BESIII current interests.
Unlike other charmonium states with the same quantum numbers and in the same
mass region, such as the $\psithr$, $\psifou$, and $\psifiv$, the $\y$ state does not have a natural place within the quark model of charmonium.
Furthermore, while being well above the $\ddb$ threshold, the $\y$ shows strong coupling to the $\pcpcjpsi$ final state,
but relatively small coupling to open charm decay modes.
These properties perhaps indicate that the $\y$ state is not a conventional state of charmonium.

Using 525~pb$^{-1}$ data collected with the BESIII detector at a CM energy of $4.260$~GeV,
$\rm BESIII$ studies the process $e^+e^-\to\pi^+\pi^-J/\psi$~\cite{bes3_zc}.
The cross section is measured to be $(62.9\pm 1.9 \pm 3.7)$~pb.
In addition, a structure (denoted as $\zc$) with a mass of $(3899.0\pm 3.6\pm 4.9)~{\rm MeV}/c^2$ and a width of
$(46\pm 10\pm 20)$~MeV is observed in the $\pi^\pm \jpsi$ mass spectrum, as shown in Figure~\ref{Zc3900-X3872}(a).
This has also been observed by Belle~\cite{belle_zc} and confirmed with CLEO data at a CM
energy of 4.17~GeV~\cite{seth_zc}.
This structure couples to charmonium and has an electric charge,
which is suggestive of a state containing more quarks than just a charm and anti-charm quark.

\section{Observation of $e^+e^- \to \gamma X(3872)$ (preliminary)}

The $X(3872)$ state was first observed by Belle~\cite{belle_x} in $B^\pm\to K^\pm\pi^+\pi^-\jpsi$ and was subsequently confirmed
by several other experiments~\cite{cdf_x,dz_x,babar_x}. Since its discovery, X(3872) has stimulated special interest for its
nature. Both BABAR and Belle have observed $X(3872) \to \gamma \jpsi$ decay, which supports $X(3872)$ being a C-even
state~\cite{babar_gjpsi,belle_gjpsi}.

\begin{figure}[htb]
\begin{center}
\includegraphics[width=0.99\textwidth]{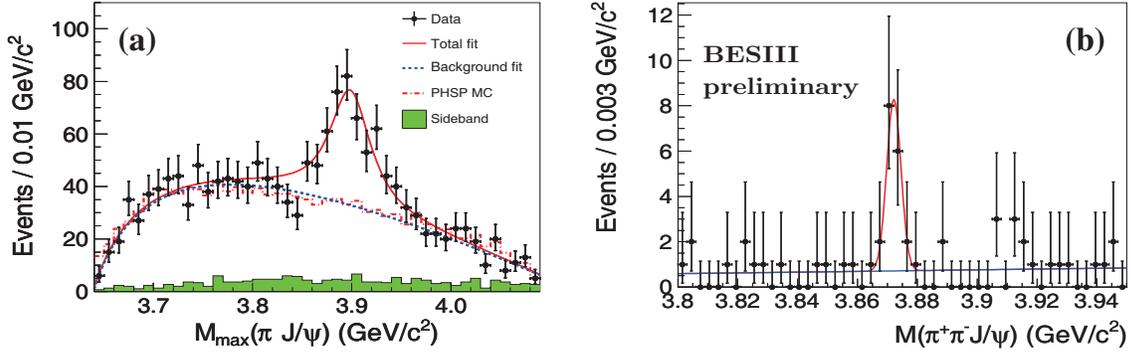} 
\caption{(a): Fit to the $M_{\rm max}(\pi^\pm\jpsi)$ distribution of $e^+e^-\to\pi^+\pi^- J/\psi$.
(b): Fit to the $M(\pi^+\pi^-\jpsi)$ distribution of $e^+e^-\to\gamma\pi^+\pi^-J/\psi$.
Dots with error bars are data, the red solid curve shows the total fit.
In (a), the blue dotted curve the shows background from the fit; the red dot-dashed histogram shows the result of a phase space MC simulation;
and the green shaded histogram shows the normalized $\jpsi$ sideband events.
In (b), the blue curve shows the background contribution.}
\label{Zc3900-X3872}
\end{center}
\end{figure}

Using the data collected with the BESIII detector at CM energies from
$4.001$~GeV to $4.420$~GeV, BESIII observes  $e^+e^- \to \gamma
X(3872) \to \gamma \pi^+\pi^-\jpsi, \jpsi \to l^+l^-(l^+l^-=e^+e^-$ or
$\mu^+\mu^-)$ for the first time. Figure~\ref{Zc3900-X3872}(b) shows the fit results to the $X(3872)$ signal.
The measured mass of the $X(3872)$, $M(X(3872))=(3872.1\pm0.8\pm0.3)$ MeV$/c^2$, agree well with previous measurements~\cite{PDG}.
The statistical significance of $X(3872)$ is 5.3 $\sigma$.
The production rate $\sigma^B[e^+e^- \to \gamma X(3872)]\times \BR[X(3872)
\to \pi^+\pi^-\jpsi]$ is measured to be $(0.32\pm0.15\pm0.02)$ pb at $\sqrt{s} =4.229$ GeV,
$(0.35\pm0.12\pm0.02)$ pb at $\sqrt{s} = 4.26$ GeV, $<0.13$ pb at $\sqrt{s} = 4.009$ GeV,
and $<0.39$ pb at $\sqrt{s} = 4.36$ GeV at the 90\% C.L.

\section{Observation of $\zcp$ in $\EE\to\pp\hc$}

The $\zc$, observed in $\EE\to \pcpcjpsi$, may couple to $\pi^\pm\hc$ and thus can be searched for in $\EE\to \pphc$.
This final state has been studied by CLEO~\cite{cleo_pphc} at CM energies from $4.000$ to    
$4.260$~GeV, and a hint of a rising cross section at $4.26$~GeV has  
been observed. An improved measurement may shed light on understanding the nature of the $Y(4260)$ as well.

BESIII studies the process $\EE\to \pphc$~\cite{bes_pphc} at 13 CM energies from
3.9000 to 4.420~GeV using data samples collected with the BESIII detector,
and are listed in Table~\ref{scan-data}.
The Born cross sections are measured at 13 energies, and are found to be of
the same order of magnitude as those of $\EE\to \pcpcjpsi$ but with a different line shape.
A narrow structure very close to the $(D^\ast\bar{D}^\ast)^\pm$ threshold (referred to as $\zcp$) with a mass of
$(4022.9\pm 0.8\pm 2.7)~{\rm MeV}/c^2$ and a width of $(7.9\pm   
2.7\pm 2.6)$~MeV is observed in the $\pi^\pm \hc$ mass spectrum, as shown in
Figure~\ref{Zc4020or4025}(a). 
This structure couples to charmonium and has an electric charge, 
which is suggestive of a state containing more quarks than just a
charm and an anti-charm quark.
We do not find a significant signal for  $\zc\to\pi^\pm\hc$ and the
production cross section is found to be smaller than 11~pb at the 90\% C.L. at 4.26~GeV.

\section{Observation of $\zcpp$ in $e^+e^- \to (D^{*} \bar{D}^{*})^{\pm} \pi^\mp$}

\begin{figure} 
\begin{center}
\includegraphics[width=0.99\textwidth]{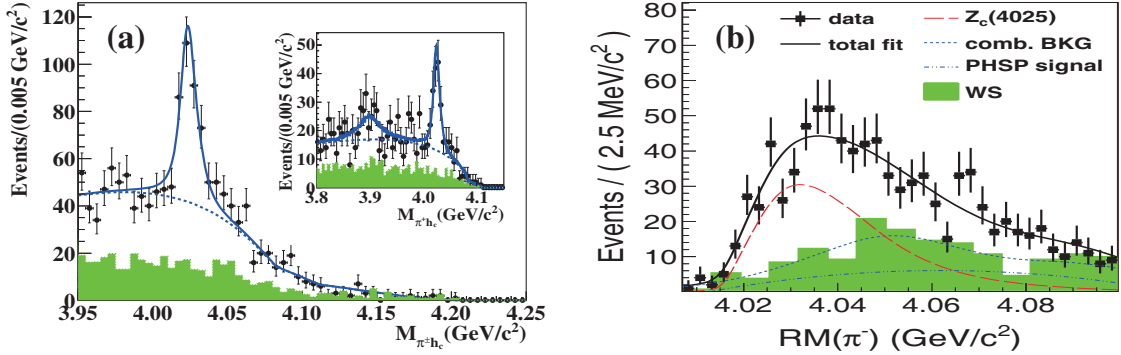}
\caption{(a): Sum of the simultaneous fits to the $M_{\pi^\pm\hc}$ distributions of $e^+e^-\to\pi^+\pi^- J/\psi$
at 4.23~GeV, 4.26~GeV, and 4.36~GeV; the inset shows the sum
of the simultaneous fits to the $M_{\pi^+\hc}$ distributions at 4.23~GeV and 4.26~GeV with $\zc$ and $\zcp$.
(b): Fit to the $\pim$ recoil mass spectrum of $e^+e^- \to (D^{*} \bar{D}^{*})^{\pm} \pi^\mp$ at 4.26~GeV.
Dots with error bars are data; the solid curves are the total fit.}
\label{Zc4020or4025}
\end{center}
\end{figure}

The mass of the $Z_c(3900)$ is about $20$~MeV higher than the $D\dstbar$ mass
threshold. Therefore, a search of $Z_c$ candidates via their direct decays into
$\dst\dstbar$ pairs is strongly motivated.
BESIII studies the process $e^+e^- \to (D^{*} \bar{D}^{*})^{\pm} \pi^\mp$~\cite{bes_pDstDst} at
a CM energy of 4.26\,GeV using a 827\,pb$^{-1}$ data sample
obtained with the BESIII detector.
Based on a partial reconstruction technique, the Born cross section is
measured to be  $(137\pm9\pm15)$\,pb.
A structure near the $(D^{*} \bar{D}^{*})^{\pm}$ threshold (referred to as $\zcpp$) in the
$\pi^\mp$ recoil mass spectrum is observed, as shown in Figure~\ref{Zc4020or4025}(b).
The measured mass and width of the structure are
$(4026.3\pm2.6\pm3.7)$\,MeV/$c^{2}$ and $(24.8\pm5.6\pm7.7)$\,MeV, respectively.
Its production ratio $\frac{\sigma(e^+e^-\to Z^{\pm}_c(4025)\pi^\mp \to (D^{*} \bar{D}^{*})^{\pm}
\pi^\mp)}{\sigma(e^+e^-\to (D^{*} \bar{D}^{*})^{\pm} \pi^\mp)}$ is determined to be $0.65\pm0.09\pm0.06$.

\section{Partial wave analysis of $J/\psi \to \gamma \eta \eta$}

Radiative $J/\psi$ decay is a gluon-rich process and
has long been regarded as one of the most promising hunting grounds for
glueballs. In particular, for a $J/\psi$ radiative decay to two
pseudoscalar mesons, it offers a very clean laboratory to search for
scalar and tensor glueballs because only intermediate states with $J^{PC}=even^{++}$ are possible.

Using 225 million $J/\psi$ events collected with the BESIII detector,
a partial wave analysis (PWA) on $J/\psi\to\gamma\eta\eta$~\cite{bes3_getaeta} was performed using the relativistic
covariant tensor amplitude method, and the results
are summarized in Table~\ref{mwb}. The scalar contributions are mainly
from $f_{0}(1500)$, $f_{0}(1710)$ and $f_{0}(2100)$, while no evident 
contributions from $f_{0}(1370)$ and $f_0(1790)$ are seen.  Recently, 
the production rate of the pure gauge scalar glueball in $J/\psi$
radiative decays predicted by the lattice QCD~\cite{Gui:2012gx} was
found to be compatible with the production rate of $J/\psi$ radiative
decays to $f_{0}(1710)$; this suggests that $f_{0}(1710)$ has a larger
overlap with the glueball compared to other glueball candidates
(eg. $f_{0}(1500)$). 

\begin {table*}[htb]
\begin {center}
\caption {Summary of the PWA results, including the masses and widths for
resonances, branching ratios of
$J/\psi\to\gamma$X, as well as the significance. The first errors are
statistical and the second ones are systematic.}
\begin {tabular}{ccccc}
\hline\hline Resonance  &Mass(MeV/$c^{2}$) &Width(MeV/$c^{2}$)
&$\BR{(J/\psi\to\gamma X\to\gamma \eta\eta)}$ &Significance\\ \hline

$f_{0}(1500)$  &1468$^{+14+23}_{-15-74}$  &136$^{+41+28}_{-26-100}$
&$(1.65^{+0.26+0.51}_{-0.31-1.40})\times10^{-5}$  &8.2~$\sigma$   \\

$f_{0}(1710)$  &1759$\pm6^{+14}_{-25}$    &172$\pm10^{+32}_{-16}$
&$(2.35^{+0.13+1.24}_{-0.11-0.74})\times10^{-4}$  &25.0~$\sigma$  \\

$f_{0}(2100)$  &2081$\pm13^{+24}_{-36}$   &273$^{+27+70}_{-24-23}$
&$(1.13^{+0.09+0.64}_{-0.10-0.28})\times10^{-4}$  &13.9~$\sigma$  \\

$f_{2}^{'}(1525)$  &1513$\pm5^{+4}_{-10}$  &75$^{+12+16}_{-10-8}$
&$(3.42^{+0.43+1.37}_{-0.51-1.30})\times10^{-5}$  &11.0~$\sigma$  \\

$f_{2}(1810)$  &1822$^{+29+66}_{-24-57}$  &229$^{+52+88}_{-42-155}$
&$(5.40^{+0.60+3.42}_{-0.67-2.35})\times10^{-5}$  &6.4~$\sigma$   \\

$f_{2}(2340)$  &2362$^{+31+140}_{-30-63}$  &334$^{+62+165}_{-54-100}$
&$(5.60^{+0.62+2.37}_{-0.65-2.07})\times10^{-5}$  &7.6~$\sigma$   \\\hline
\hline

\end {tabular}  
\label{mwb}
\end {center} 
\end {table*}

\section{Measurement of $D^+ \to \mu^+ \nu$ (preliminary)}

The decay rate is proportional $f_D^2$, here $f_D$ is the $D^+$ decay
constant. By measuring the branching fraction of $D^+ \to \mu^+ \nu$, 
the $f_D$ can be extracted with external input of $V_{cd}$. 

\begin{figure}[htb]
\centering
\includegraphics[width=0.75\textwidth]{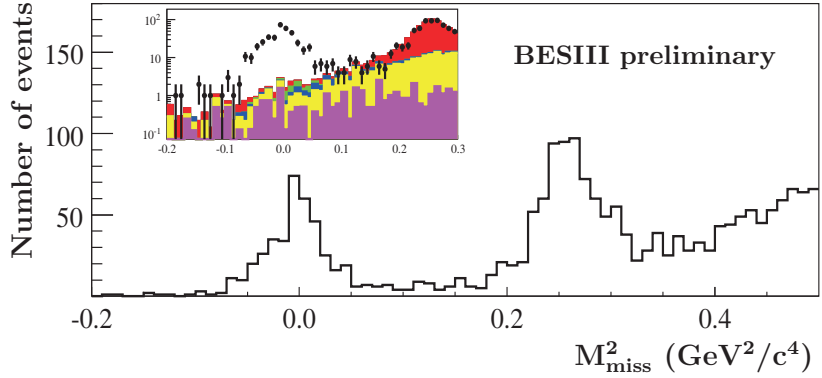}
\caption{$M_{miss}^2$ distribution.  
Inset: log plots of $M_{miss}^2$ with stacked backgrounds in color (non-$D\bar{D}$ events in magenta, other $D\bar{D}$ in yellow, 
$D^+ \to \tau^+ \nu$ in blue, $D^+ \to \pi^+\pi^0$ in green, and $D+ \to K_L \pi^+$ in red ).}
\label{fig:munu}
\end{figure}

BESIII extracts the $f_D$ from the leptonic decays $D^+ \to \mu^+ \nu$~\cite{bes_munu}
using $2.92$~fb$^{-1}$ data taken at $\sqrt{s}=3.773$~GeV.
The $D^+$ meson are produced from $\psi(3770) \to D^+D^-$.
The $D^-$ mesons are reconstructed in nine non-leptonic decay modes.
The signal for $D^+ \to \mu^+ \nu$ is observed in the distribution of $M_{miss}^2 = E_{miss}^2 - p_{miss}^2$, 
where $E_{miss}$ and $p_{miss}$ are the missing energy and momentum due to the undetectable neutrino in the detector.
The $M_{miss}^2$ distribution is presented in Figure~\ref{fig:munu}, where a remarkably clean signal peak
at zero is evident.

We observe a signal of $377.3 \pm 20.6 \pm 2.6$ events above a background
of 47.7 events.  From this signal, we extract:
${\cal B}(D^+ \to \mu^+ \nu) \,=\, (3.74 \pm 0.21 \pm 0.06) \times 10^{-4}$
and $f_D \,=\, (203.91 \pm 5.72 \pm 1.97) \;{\rm  MeV} \;.$
This is more precise than the previous best measurement of $f_D \,=\, (205.8 \pm 8.5 \pm 2.5)$ MeV,
based on 818 pb$^{-1}$ from CLEO-c\cite{fD_CLEO}.

\section{Measurement of $D^0 \to K^- e^+\nu$ and $D^0 \to \pi^- e^+\nu$ (preliminary)}

BESIII extracts the form-factors $f_{\pi,K}(q^2)$
from the semileptonic decays $D^0 \to K^- e^+ \nu$ and $D^0 \to \pi^- e^+ \nu$\cite{bes_kenupienu}
using one-third of $2.92$~fb$^{-1}$ data taken at $\sqrt{s}=3.773$~GeV.
Here, $q^2 = m_{e\nu}^2$ and these form factors describe the effects of 
meson structure in the decay, relative to idealized free-quark decay.   

\begin{figure}[htbp] 
\begin{center}
\includegraphics[width=0.75\textwidth]{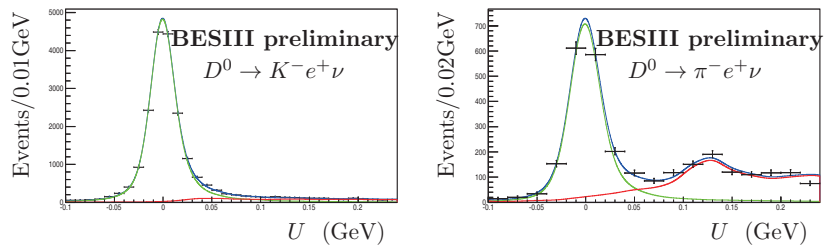}
\caption{$U \,=\, E_{miss} - p_{miss}$ distributions for
$D^0 \to K^-e^+\nu$ (left), $D^0 \to \pi^-e^+\nu$ (right).    
The blue total fit curve is the sum of a green signal shape and
a red background term.}
\label{Umiss} 
\end{center}
\end{figure}

The $D^0$ meson are produced from $\psi(3770) \to D^0\bar{D}^0$.
The $D^0$ mesons are reconstructed in four non-leptonic decay modes.
The amount of signal events is determined by fitting the distribution of 
$U = E_{miss} - p_{miss}$; the ``miss'' quantities, representing the unobserved neutrino,
are analogous to those in the previous analysis.  
For signal, $U$ peaks at zero and is similar to a missing-mass-squared.
Fits to the $U$ distributions in Fig. \ref{Umiss}
lead to the branching fraction results: $\BR(D^0 \to   K^-e^+\nu) = (3.542 \pm 0.030 \pm
0.067)$\% and $\BR(D^0 \to \pi^-e^+\nu) = (0.288 \pm 0.008 \pm 0.005$)\%,
which are consistent with the previous measurement from CLEO~\cite{FF_CLEO}.
The form-factor analysis is provided in Ref.~\cite{bes_kenupienu}.

\section{Summary}

The BESIII experiment has collected the world's largest samples of $\jpsi$, $\psi(2S)$, $\psi(3770)$, $\psi(4040)$, $Y(4260)$ and $Y(4360)$ decays.
Based on these samples, BESIII has produced a large amount of results on the searches for $XYZ$ states,
charmonium spectroscopy and decays, light hadron spectroscopy, and $D$ meson decays.
The data taking at high luminosities will go on for years. Many new discoveries and precision measurements
are expected to be coming soon.

\end{document}